\documentclass[aps,prl,superscriptaddress,showpacs,showkeys]{revtex4-1}

\usepackage{amsmath}
\usepackage{graphicx}
\usepackage{color}\definecolor{RED}{rgb}{1,0,0}\definecolor{BLUE}{rgb}{0,0,1}
\usepackage[normalem]{ulem}

\begin{document}


\title{General Initial Value Problem for the Nonlinear Shallow Water Equations: Runup of Long Waves on Sloping Beaches and Bays}



\date{Authors are arranged in the alphabetical order}

\author{Dmitry Nicolsky}
\affiliation{Geophysical Institute, University of Alaska Fairbanks, USA}

\author{Efim Pelinovsky}
\affiliation{Nizhny Novgorod State Technical University n.a. R. Alekseev, Russia}
\affiliation{Special Research Bureau for Automation of Marine Researches, Yuzhno-Sakhalinsk, Russia.}
\affiliation{Institute of Applied Physics, Nizhny Novgorod, Russia}

\author{Amir Raz}
\affiliation{Geophysical Institute, University of Alaska Fairbanks, USA}
\affiliation{Department of Mathematics and Statistics, University of Alaska Fairbanks, USA}

\author{Alexei Rybkin}
\affiliation{Department of Mathematics and Statistics, University of Alaska Fairbanks, USA}


\begin{abstract}
We formulate a new approach to solving the initial value problem of the shallow water-wave equations utilizing the famous Carrier-Greenspan transformation [G. Carrier and H. Greenspan, J. Fluid Mech. 01, 97 (1957)]. We use a Taylor series approximation to deal with the difficulty associated with the initial conditions given on a curve in the transformed space. This extends earlier solutions to waves with near shore initial conditions, large initial velocities, and in more complex U-shaped bathymetries; and allows verification of tsunami wave inundation models in a more realistic 2-D setting\footnote{To appear in Physics Letters A.}.
\end{abstract}


\maketitle


\section{Introduction}

Tsunami modeling and forecast is an important scientific problem impacting coastal communities worldwide. Many models for tsunami wave propagation use the 2+1 shallow water equations (SWE), an approximation of the Navier-Stokes equation \cite{Kanoglu15}. These numerical  models must be continuously verified and validated to ensure the safety of coastal communities and infrastructure \cite{NTHMP12}. Apart from verification against data from actual tsunami events, numerical models are also extensively verified against analytical solutions of the 2+1 SWE which exist for idealized bathymetries \cite{Synolakis08}. These analytical solutions also give important qualitative insight to tsunami run-up and amplification.

Typically, the process of tsunami generation is considered as an instant vertical motion of the sea bottom ignoring the water velocities in the source. However, incorporation of the water velocities into the initial conditions is important from physical point of view, see for instance \cite{Kanoglu06}. For a more complete analysis of tsunami hydrodynamics, modeling and forecast, we refer the reader to \cite{Synolakis06,Synolakis08,Kanoglu15,NTHMP12,TsunamiDynamics15}.

A classical example of an analytical solution for the 2+1 SWE is computing the run-up of long-waves on a sloping beach \cite{Synolakis87}. Because of the symmetric bathymetry, the 2+1 SWE reduce to the 1+1 SWE, which could be solved directly in the physical space \cite{Antuono10} or in the new coordinates using the Carrier-Greenspan transformation \cite{Carrier1}. The SWE in the transformed coordinates has been extensively studied as an initial value problem (IVP) \cite{Yeh03,KANOGLU04,Kanoglu06,Harris15,Aydin17} and as a boundary value problem \cite{Synolakis87,Antuono07}. The IVP for waves with non-zero initial velocity have been previously derived using a Green's function in \cite{Yeh03,Kanoglu06}, though both solutions imply assumptions regarding the initial velocity as discussed later. Thus, the complete and exact solution to the IVP for waves with nonzero initial velocities remains a long standing open problem \cite{Tuck72,Kanoglu06,Antuono07}.

The 1+1 SWE for the sloping beach have recently been generalized to model waves in sloping narrow channels using the cross-sectionally averaged 1+1 SWE \cite{Zahibo05}. Furthermore, the hodograph transform given by \cite{Carrier1} can be generalized to sloping bays with arbitrary cross sections, allowing a much richer problem to study \cite{Rybkin}. Though the cross-sectionally averaged 1+1 SWE have no analytical solution for bays with arbitrary cross sections, an analytical solution exists for symmetric U-shaped bays, i.e bays with a cross section $z\propto y^m$ \cite{Zahibo05,Rybkin,Anderson}. The known solution for sloping beaches is an asymptotic solution of such bays when $m\rightarrow\infty$.

In this letter we propose a new approach to solve the IVP for the cross-sectionally averaged 1+1 SWE in U-shaped bays for waves with arbitrary initial velocities exactly. Our solution uses a Taylor expansion to deal with the initial data given on a curve (under the Carrier-Greenspan transformation the line $t=0$ is mapped to a curve in the transformed plane), a problem that was not sufficiently treated in the previous solutions. This allows run-up computation of near shore long waves, unlike the previous IVP solutions that require the initial wave to be far from shore \cite{Kanoglu06}. Additionally, we present some qualitative geophysical implications using this new solution.

\section{Solution of the IVP}

The cross-sectionally averaged 1+1 SWE for U-shaped bays describe the evolution of long waves in a sloping narrow bay with an unperturbed  water height $h(x)= x$ along the main axis of the bay in dimensionless form. The wave is assumed to propagate uniformly through the bay in the $x$ direction, a valid assumption as shown in \cite{Shimozono16,Anderson}. The cross-sectionally averaged SWE for such bays in dimensionless  form are given by \cite{Zahibo05,Anderson} to be
\begin{subequations}
\label{eq:CSA SWE}
\begin{eqnarray}
\eta_t+u(1+&\eta_x&)+\beta^2(x+\eta)u_x = 0, \\
u_t&+&uu_x+\eta_x=0,
\end{eqnarray}
\end{subequations}
where $u(x,t)$ and $\eta(x,t)$ are the horizontal depth-averaged velocity and free-surface elevation along the main axis of the bay, respectively, and $\beta^2=m/(m+1)$ is the wave propagation speed along a constant depth channel. An arbitrary scaling parameter $l$ is used to introduce the dimensionless variables $x=\tilde{x}/l$, $\eta=\tilde{\eta}/(l\alpha)$, $u=\tilde{u}/\sqrt{g\alpha l}$ and $t=\tilde{t} \sqrt{g\alpha/l}$. Here $\tilde{x},\tilde{t},\tilde{\eta}$ and $\tilde{u}$ are the dimensional variables, $g$ is the gravitational acceleration, and $\alpha$ is the slope of the incline.

\begin{figure}
	\includegraphics[width=0.89\textwidth]{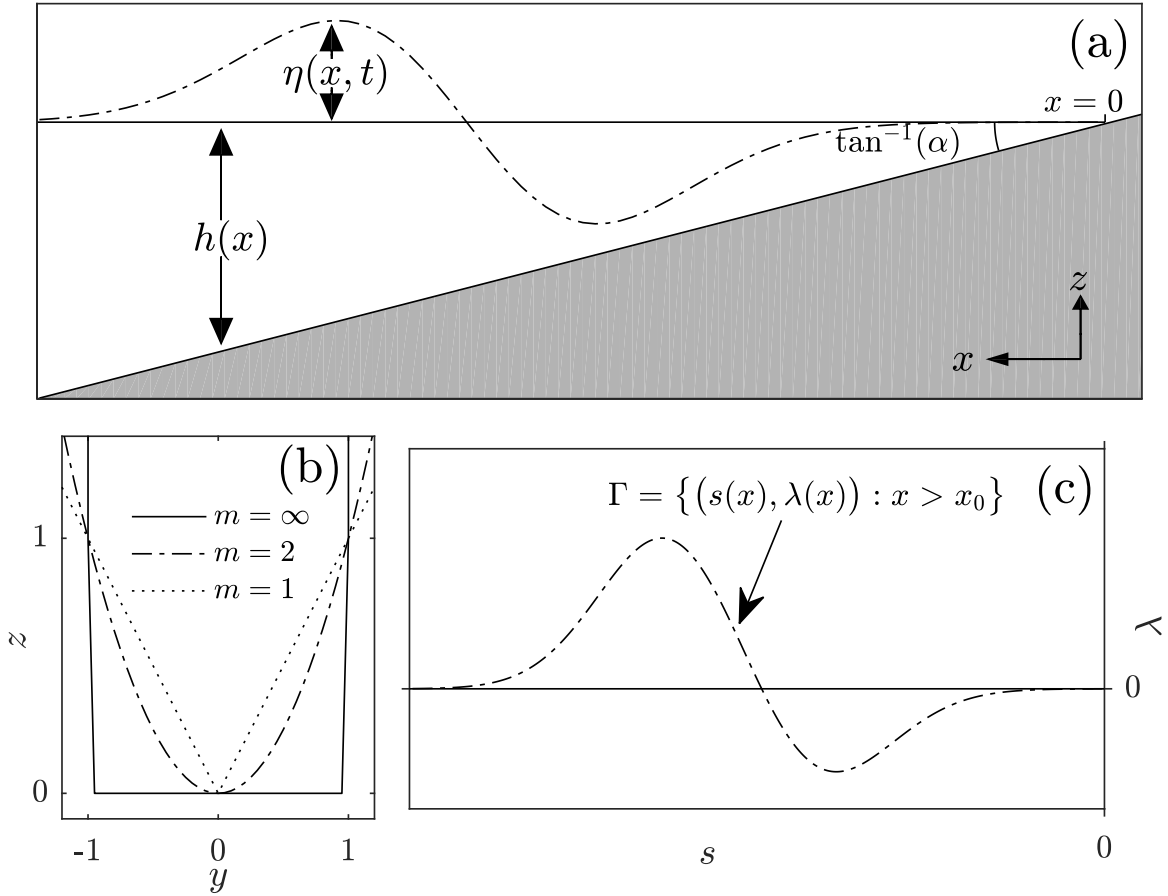}
    \caption{Definition sketch: (a) $x-z$ cross section with perturbed and unperturbed water heights, not to scale. (b) $y-z$ cross section of a plane beach where $m=\infty$, a parabolic bay where $m=2$, and a V-shaped bay where $m=1$. (c) The curve $\Gamma$ on which the initial conditions are prescribed in the transformed space for the wave shown in (a), not to scale.}
    \label{fig:intro}
\end{figure}

We use the form of the Carrier-Greenspan transformation presented in \cite{Tuck72},
\begin{subequations} \label{eq:CG}
\begin{eqnarray}
\varphi   =  u, \ & \ \ \ \ \ &  \psi=\eta+\frac{u^{2}}{2}%
\label{eq:CG1}\\
s  =  x  + & \eta, \ \ \ \ \ \ \ \ \ &  \lambda=t-u, \label{CG 2}%
\end{eqnarray}
\end{subequations}
to reduces (\ref{eq:CSA SWE}) to the linear system
\begin{equation}
\Phi_\lambda+A(s)\Phi_s+B\Phi = 0,
\label{eq:lin sys}
\end{equation}
where $\Phi(s,\lambda)=\begin{pmatrix} \varphi(s,\lambda) \\ \psi(s,\lambda)\end{pmatrix}$, $A(s)=\begin{pmatrix}0&1\\ \beta^2 s& 0\end{pmatrix}$, and $B=\begin{pmatrix}0&0\\1&0\end{pmatrix}$. This form of the Carrier-Greenspan transformation has two useful properties, the moving shoreline is fixed at $s=0$ and the resulting linear system, (\ref{eq:lin sys}), is the linear SWE. For comparison to other texts, specifically \cite{Kanoglu06,Rybkin,Carrier1,Synolakis87}, the transform variable $\sigma=2\sqrt{s}/ \beta$ is typically used, along with the introduction of a potential function to form a single linear second order  partial differential equation.

We consider (\ref{eq:lin sys}) with the general initial conditions in physical space $\eta(x,0)=\eta_0(x)$ and $u(x,0)=u_0(x)$. Under transformation (\ref{CG 2}), $\eta_0(x)$ and $u_0(x)$ transform into initial conditions on a parameterized curve $\Gamma$ in the $(s,\lambda)$ plane, depicted in Fig. \ref{fig:intro}c, which leads to a non-trivial IVP. It is natural to parameterize this curve using the coordinate $x$, $\Gamma=\{\Gamma(x): x>x_0\}$, where
\begin{equation} \label{eq:gamma}
 \Gamma(x) = \big( s(x), \lambda(x) \big) =  \big(x+\eta_0(x), -u_0(x) \big),
 \end{equation}
and $x_0$ is the $x$ position of the shoreline at time $t=0$. The initial condition is then given by
\begin{equation} \label{eq:initial}
 \Phi|_{\Gamma(x)} = \Phi_0(x) = \begin{pmatrix}u_0(x) \\  \eta_0(x)+u_0^2(x)/2 \end{pmatrix}.
 \end{equation}

A general solution to (\ref{eq:lin sys}) can be found using the Hankel transform to be \cite{Zahibo05,Anderson}
\begin{subequations}\label{eq:Phi}
\begin{eqnarray}
\nonumber \psi(s,\lambda)  &  = & s^{-\frac{1}{2m}}\int_{0}^{\infty}\left\{  a(k)\cos
(\beta k\lambda)+b(k)\sin(\beta k\lambda)\right\} \\ & & \times J_{1/m}\left(  2k\sqrt{s}\right)  dk,  \label{eq:psi} \\
\nonumber \varphi(s,\lambda)  &  = & \frac{1}{\beta}s^{-\frac{1}{2m}-\frac{1}{2}}\int_{0}^{\infty
}\left\{  a(k)\sin(\beta k\lambda)-b(k)\cos(\beta k\lambda)\right\} \\
& & \times J_{1/m+1}\left(2k\sqrt{s}\right)  dk, \label{eq:phi}
\end{eqnarray}
\end{subequations}
where $J_\nu(\alpha)$ is the Bessel function of the first kind of order $\nu$, and $a(k)$ and $b(k)$ are arbitrary functions determined by the initial conditions. We note that the apparent singularities at $s=0$ are removed using the asymptotic of the Bessel function of the first kind around zero.

In the piston model of generation, i.e. with zero initial velocity, the curve $\Gamma$ coincides with the line $\lambda=0$. For arbitrary initial conditions on the line $\lambda=0$, using the inverse Hankel transform, we have that
\begin{subequations}
\label{eq:a and b}
\begin{eqnarray}
a(k) & = & 2 k \int_0^\infty \psi(s_*,0) s_*^{\frac{1}{2m}} J_{\frac{1}{m}}\left(2k\sqrt{s_*}\right) d s_*,  \\
b(k) & = & -2 \beta k \int_0^\infty \varphi(s_*,0) s_*^{\frac{1}{2m}+\frac{1}{2}} J_{\frac{1}{m}+1}\left(2k\sqrt{s_*}\right) d s_*. \ \ \ \ \ \ \ \
\end{eqnarray}
\end{subequations}
 For waves with zero initial velocity, using (\ref{eq:initial}) and a simple change of variables,  (\ref{eq:a and b}) simplifies to $b(k)=0$, and
\begin{eqnarray}
a(k) =  2 k \int_{x_0}^\infty & \eta_0(x_*) \big(x_*+\eta_0(x_*)\big)^{\frac{1}{2m}} J_{\frac{1}{m}} \left(2k\sqrt{x_*+\eta_0(x_*)}\right)  \nonumber \\
& \times(1+\eta_0'(x_*))dx_*,  \label{eq:a zero u}
\end{eqnarray}
where primes denote derivatives in $x$. Using (\ref{eq:Phi}), $\varphi(s,\lambda)$ and $\psi(s,\lambda)$ can be computed . The solution is then transformed to physical space using (\ref{eq:CG}). The solution over a large number of grid points can be found by interpolation using Delaunay triangulation, as in \cite{Anderson}. Alternatively, Newton-Raphson iterations can be used to find the solution for a particular location $x$ or time $t$, as in \cite{Synolakis87,KANOGLU04}.

If the initial wave has an initial velocity, the curve $\Gamma$ may be complicated so that an exact solution does not exist. Reference \cite{Kanoglu06} used the approximation $s\approx x$ to find an approximate solution for wave run-up on plane beaches using a Green's function. This solution is similar  to (\ref{eq:Phi}) and (\ref{eq:a and b}) under the appropriate change of variables and order of integration, with $m=\infty$. Although \cite{Kanoglu06} defined the initial condition along a curve in the transform space, the obtained results are expressed as trigonometric functions of $\lambda-\lambda_0(s)$, where $\lambda_0(s)=-u_0(s)$. Therefore, the resulting function solves the governing partial differential equation approximately as long as ${\partial u_0}/{\partial x}\approx 0$. An interested reader could find further details in the Appendix. Even though approximation in \cite{Kanoglu06,Yeh03} are applicable to many geophysical conditions, i.e. when ${\partial u_0}/{\partial x}\approx 0$ or $u_0 \approx 0$, but for near shore waves with large initial velocities those solutions might break down.

To overcome this difficulty, we propose to project the initial conditions onto the line $\lambda=0$ using Taylor's theorem. For such a projection to exist, the transformation $s=x+\eta_0(x)$ must be bijective, and therefore $\eta_0'(x)>-1$ for all $x$. We will call the projection of the initial condition to $n^{\text{th}}$ order
\begin{equation} \label{eq:taylor}
\Phi_n(x) = \Phi_n(s(x)) =\sum_{k=0}^n \left[\frac{(-\lambda)^k}{k!}\frac{\partial^k \Phi}{\partial\lambda^k}\right]_{\Gamma(x)}.
\end{equation}
Once the desired $\Phi_n(x)$ is obtained, $a(k)$ and $b(k)$ can be computed using (\ref{eq:a and b}). Furthermore, a simple change of variables $s_*=x_*+\eta_0(x_*)$ in the integration of $a(k)$ and $b(k)$, similar to the change of variables in (\ref{eq:a zero u}) for waves with zero initial velocity, nullifies the need for the approximation $s\approx x$ required in the previous IVP solutions in \cite{Kanoglu06,Yeh03}. The complete solution can then be found using (\ref{eq:CG}) as described above. This method allows computing the solution to any desired accuracy without assumptions on the initial velocity profile.

The partial derivatives in (\ref{eq:taylor}) are not explicitly computable from our initial conditions. To put (\ref{eq:taylor}) in explicit form we use the chain rule
\begin{eqnarray}\label{eq:chain} \nonumber
\Phi_0' = \frac{d\Phi_0}{dx}&=& (1+\eta_0') \frac{\partial \Phi}{\partial s}\Big|_{\Gamma} - u_0' \frac{\partial\Phi}{\partial\lambda} \Big|_{\Gamma}\\
\nonumber
&=& (1+\eta_0') \frac{\partial \Phi}{\partial s}\Big|_{\Gamma} + u_0' \left[A\frac{\partial \Phi}{\partial s}+B\Phi\right]_{\Gamma}\\
&=&D\frac{\partial \Phi}{\partial s}\Big|_{\Gamma}+u_0'B\Phi\Big|_{\Gamma},
\end{eqnarray}
where $D= (1+\eta_0')I+u_0'A $, and $I$ is the $2$-by-$2$ unit matrix. Noting that
\[
\Phi_1 = \left[\Phi- \lambda\frac{\partial\Phi}{\partial\lambda}\right]_\Gamma=\Phi|_\Gamma+u_0\frac{\partial \Phi}{\partial \lambda}\Big|_{\Gamma} = \Phi|_\Gamma-u_0\left[A\frac{\partial \Phi}{\partial s}+B\Phi\right]_{\Gamma}
\]
and after substituting $\Phi_s$ from (\ref{eq:chain}), while recalling that $\Phi_0=\Phi|_{\Gamma}$, we obtain
\[\Phi_1 = \Phi_0+u_0 \left\{u_0' A D^{-1} B\Phi_{0}-B \Phi_{0}- AD^{-1} \Phi_{0}' \right\}.\]
Similarly, higher order terms can be found using the recursive relationship
\begin{eqnarray}\label{ref:iterative_phi}
\Phi_n &= & \Phi_{n-1} + \frac{(u_0)^n}{n!} \left[ u_0'A D^{-1} B-B - AD^{-1}\frac{d}{dx} \right]^n \Phi_0 \nonumber\\
&= &\Phi_{n-1} + \frac{1}{n} \left[  u_0 u_0'A D^{-1}B-u_0 B+(n-1)u_0'AD^{-1} \right] \nonumber \\
& & \ \ \times (\Phi_{n-1} - \Phi_{n-2} )  - \frac{u_0}{n}AD^{-1}(\Phi_{n-1}' - \Phi_{n-2}') .
\end{eqnarray}

It is important to understand the limitations of this approach. Based on the given formulas, the projection of the initial conditions remains one-to-one so long as $D$ remains nonsingular, which corresponds to $\left(1+\eta_0'\right)^2 - (u_0')^2\beta^2 (x+\eta_0) > 0$ for all $x$. To simplify this condition, we define the breaking parameter to be
$$\text{Br} = \max_{x\in [x_0,\infty)} \left\{ \beta |u_0'(x)| \sqrt{x+\eta_0(x)} - \eta_0'(x) \right\} .$$
So long as $\text{Br}<1$ the projection is one-to-one. We emphasize that our solution method is only valid for initial profiles and velocities that satisfy the conditions $\text{Br}<1$ and $\eta_0'>-1$. From (\ref{CG 2}) and (\ref{eq:CSA SWE}), it follows that the condition $\text{Br}<1$ is equivalent to the Jacobian of the Carrier-Greenspan transform not vanishing at $t=0$. For application purposes, typical tsunami waves have a  much larger wavelength than wave height \cite{Kanoglu06}, and thus satisfy $|u_0'|,|\eta_0'|\ll 1$. Therefore these conditions pose little restrictions for the modeling of geophysical long waves (i.e. ``localized'' water disturbance such as Gaussian, solitary or N-waves having the water level $\eta_0(x)$ and velocity $u_0(x)$ infinitesimal for $x\to\infty$).

\section{Qualitative Analysis}

\begin{figure}
	\includegraphics[width=0.8\textwidth]{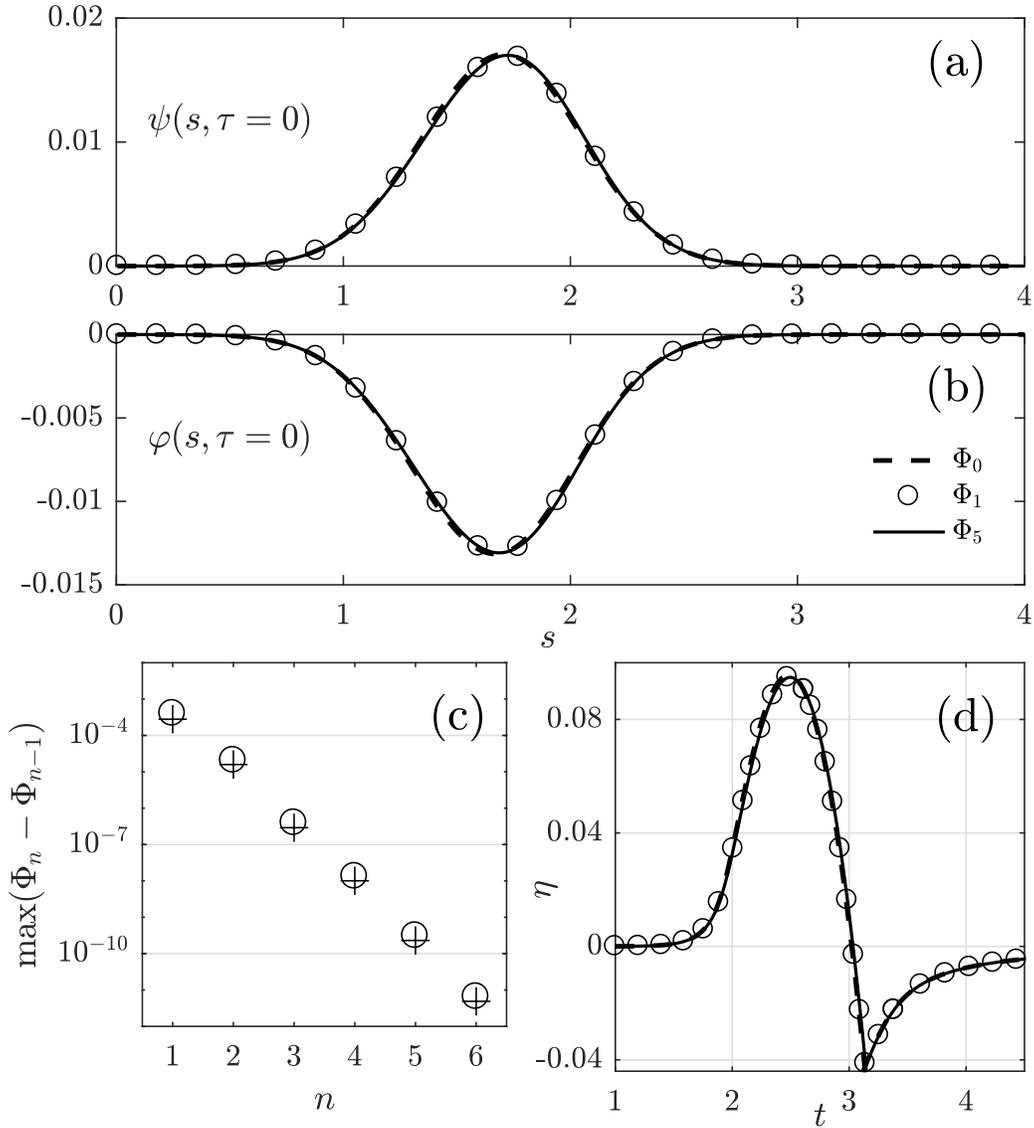}
    \caption{The transformation of an incident Gaussian wave given by (\ref{eq:gauss}) with $a=0.017$, $b=4.0$ and $x_0=1.69$. (a) $\psi(s, \lambda=0)$ to zeroth, first, and fifth order. (b) $\varphi(s, \lambda=0)$ to zeroth, first, and fifth order. (c) The maximum difference between approximations of the initial profile: $\circ$ and $+$ represent the maximum difference in $\varphi(s,\lambda=0)$and $\psi(s,\lambda=0)$, respectively. (d) Comparison of the shoreline displacement for the three orders of approximation ($n=0, 1, 5$) of projection of the initial conditions in (a) and (b), the line code is the same as in plot (b).}
    \label{fig:initial profile}
\end{figure}
Although use of solitary waves as proxies for geophysical tsunamis is subject to the discussion \cite{MadsenFuhrman}, we seek to validate our proposed solution to the IVP by checking the convergence of the solution and comparing it to the previous solution in \cite{Kanoglu06}. Reference \cite{Kanoglu06} analyzed a Gaussian initial wave profile defined by
\begin{equation}\label{eq:gauss}
\eta_0(x) = ae^{-b(x-x_0)^2},
\end{equation}
with the linear approximation of initial velocity $u_0(x)=-\eta_0(x)/\sqrt{x}$. At the same time, the developed method (\ref{eq:taylor}) also allows computations for the nonlinear approximation to the initial velocity $u_0(x)=-2(\sqrt{x+\eta_0}-\sqrt{x})$.

In Fig. \ref{fig:initial profile}a,b we present several orders of approximations for the transformation of the initial profile of such wave. We note that the iterations converge very rapidly, with the higher order approximations overlapping with the zeroth order approximation. The maximum change in the initial profile between iterations is presented in Fig. \ref{fig:initial profile}c. The convergence appears linear, with the difference after just six iterations approaching the machine limit. In Fig. \ref{fig:initial profile}d we present the shoreline run-up for several different orders of approximation. Notice that the zeroth order profile coincides almost completely with the higher order approximations. Because of this, both previous solutions \cite{Kanoglu06,Yeh03} with assumptions ${\partial u_0(x)}/{\partial x}\approx 0$ or $u_0(x)\approx 0$, give valid results for such initial conditions. Figure \ref{fig:initial profile}d allows direct comparison to figure 3d in \cite{Kanoglu06}.

\begin{figure}
	\includegraphics[width=0.79\textwidth]{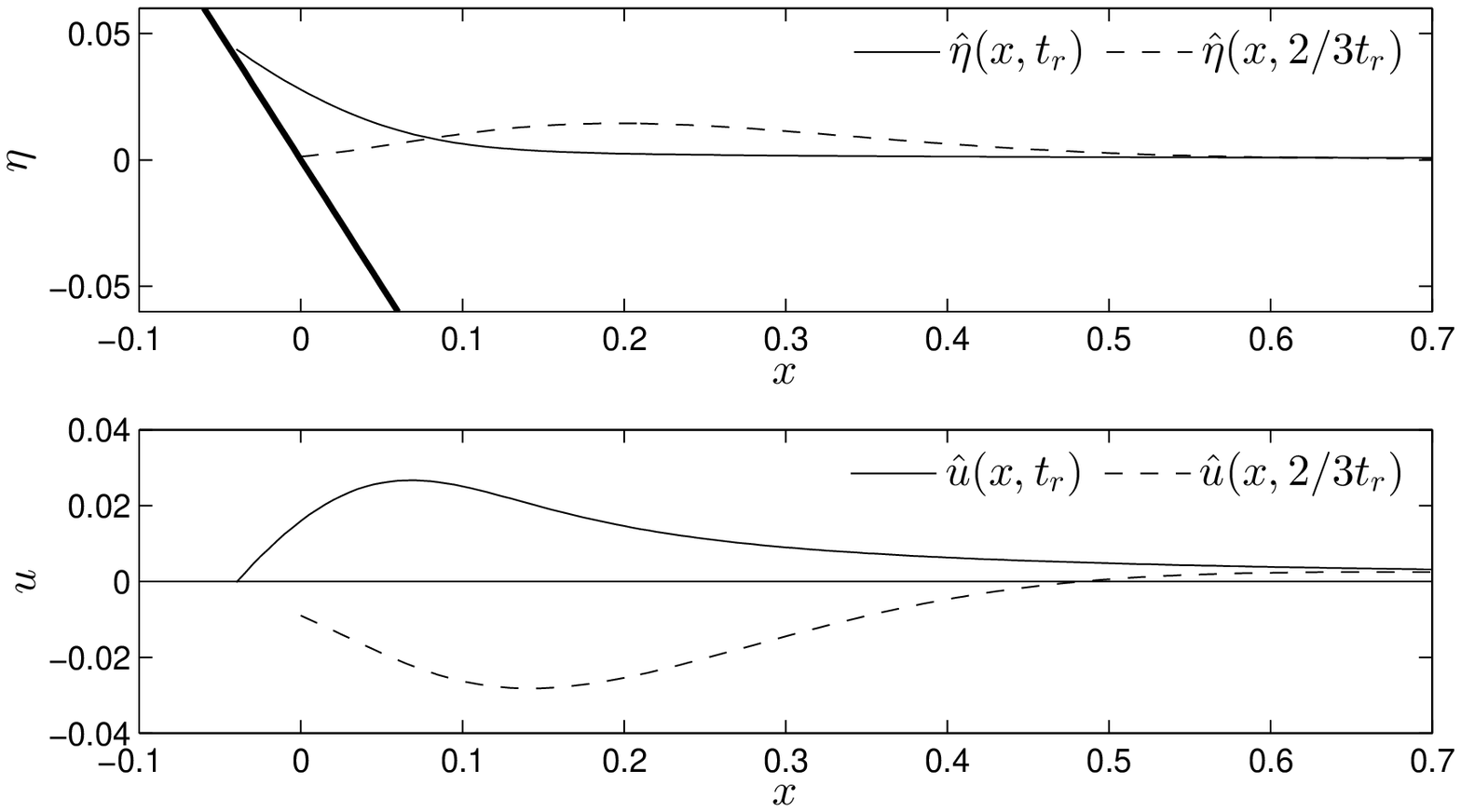}
    \caption{Profiles of the water level $\hat\eta$ (top) and water velocity $\hat{u}$ (bottom) for the initial condition: a zero-velocity Gaussian waves given by (\ref{eq:gauss}) with $a=0.017$, $b=4.0$ and $x_0=1.69$ running up a plane beach ($m=\infty$).}
    \label{fig:validation1}
\end{figure}

\begin{figure}
	\includegraphics[width=0.79\textwidth]{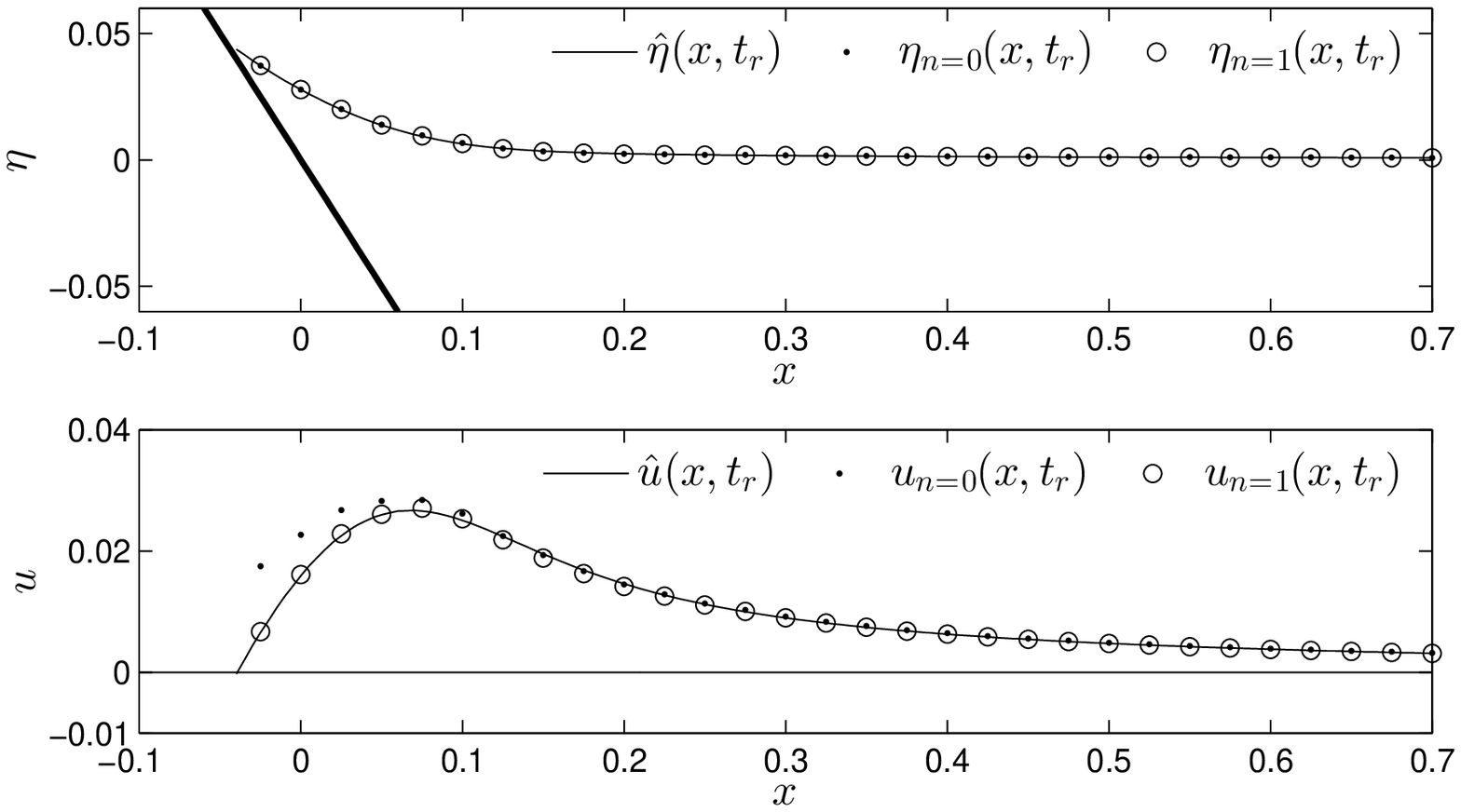}
    \caption{Comparison of the water level $\eta$ (top) and water velocity $u$ (bottom) at the time  $t=t_r$ of maximum runup. Quantities $\eta_n$ and $u_n$ are computed for different approximations of $\{\Phi_n\}_{n=0}^1$ from the transfer point $t=2/3t_r$, while taking a non-zero velocity $\hat{u}(x,2/3t_r)$.}
    \label{fig:validation2}
\end{figure}

To validate the presented approach, we consider a Gaussian-shaped initial wave ($a=0.017$ and $b=4.0$) centered at the distance of $x_0=1.69$ from the shore of a plane beach ($m=\infty$). We additionally assume that the initial water velocity $u_0$ is zero. For this case, we can easily compute water level $\hat\eta$ and velocity $\hat{u}$ until the moment of maximum runup at $t_r\approx2.475$. The obtained profiles $\hat\eta(x,t_r)$ and $\hat{u}(x,t_r)$ are later used to validate the proposed methodology to compute water dynamics from the initial conditions with a non-zero velocity as follows. In particular, while computing $\hat\eta(x,t_r)$ and $\hat{u}(x,t_r)$, we save the water level and velocity at some transient time, $t=2/3t_r$, when the wave is about to runup on the shore. Fig. \ref{fig:validation1} displays wave profile and velocity at the time of maximum runup and at the transient point.

At the transient point, we use $\hat\eta(x,2/3t_r)$ and $\hat{u}(x,2/3t_r)$ to setup initial conditions for the general initial value problem. Since the water velocity $\hat{u}(x,2/3t_r)\ne0$, we project $\hat\eta(x,2/3t_r)$ and $\hat{u}(x,2/3t_r)$ onto the line $\lambda=0$ via (\ref{eq:taylor}) and then apply formulae (\ref{eq:Phi}) to model propagation of the wave further onshore. Furthermore, we investigate an accuracy of the obtained solution $\eta_n(x,t)$ and $u_n(x,t)$ by considering different orders of approximation, $n$.

Comparison of the water level profiles $\hat\eta(x,t_r)$, $\eta_{n=0}(x,t_r)$, and $\eta_{n=1}(x,t_r)$ at the moment of maximum runup is shown in the top plot in Fig. \ref{fig:validation2}. One may note that even for the zeroth approximation, $n=0$, the match between the water elevation profiles is rather good. Comparison between the water velocities $\hat{u}(x,t_r)$ and $u_{n=0}(x,t_r)$ shows a discrepancy near the shore. However, the first order approximation for the velocity $u_{n=1}(x,t_r)$ provides a satisfactory match with the analytical solution $\hat{u}(x,t_r)$ computed for the zero-initial velocity. This comparison implies that the proposed methodology can be satisfactory applied to model wave propagation with a non-zero initial velocity.

\begin{figure}
	\includegraphics[width=0.5\textwidth]{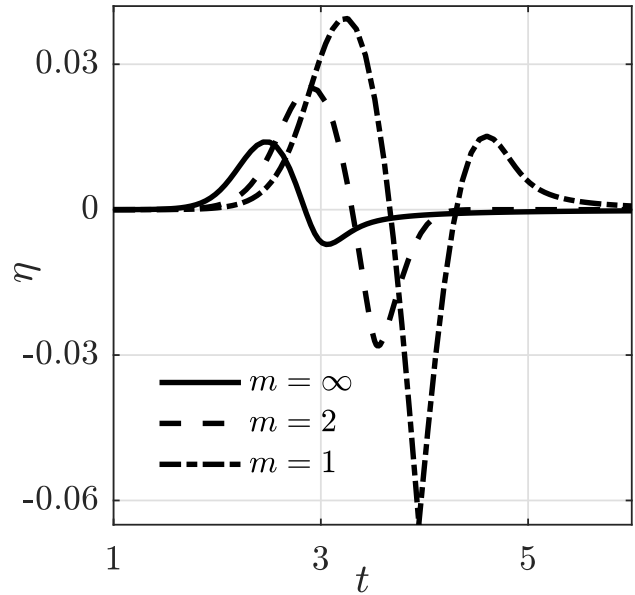}
    \caption{ Comparison of the shoreline displacement for identical incident Gaussian waves given by (\ref{eq:gauss}) with $a=0.0025$, $b=4.0$ and $x_0=1.69$, in a plane beach ($m=\infty$), parabolic bay ($m=2$), and V-shaped bay ($m=1$).}
    \label{fig:shoreline}
\end{figure}

With converges of our solution verified, we highlight some geophysical implications that our solution has. In particular, our solution allows analysis of long wave waves in 2-D U-shaped bathymetries, rather than only on plane beaches. In light of this, and because local bathymetry can significantly affect run-up height \cite{Synolokis97,Synolakis06}, we analyze the effect of the bay shape on the height of maximum run-up. We look at the run-up of the same Gaussian wave with the linear approximation of initial velocity in three different bays: a plane beach ($m=\infty$), a parabolic bay ($m=2$) and a V-shaped bay ($m=1$). The shoreline displacement of these three run-up scenarios is presented in Fig. \ref{fig:shoreline}. We see that the maximum run-up is almost twice as large in parabolic bays, and almost three times as large in V-shaped bays, than over a regular plane beach. This result shows that long waves can be greatly amplified in heads of narrow bays, and can help explain amplification of long waves in narrow channels and bays.

These findings have profound implications not only for coastal engineering in narrow bays or channels, but also for hydraulic engineering. For example, the Vajont dam in Italy is located at the head of narrow V-shaped valley. In 1963, a landslide caused a $\sim 200$ meter high wave that overcame the dam and caused massive destruction to towns downstream. Such events can be modeled using shallow water theory \cite{Vajont11}, and our results can help explain why the wave was highly amplified in the narrow channel. Understanding the effects of narrow channel bathymetry on wave amplification is crucial for ensuring the safety of communities near bays and dams.

\section{Discussion and Conclusions}

The current models used for tsunami forecast have primarily been verified against the analytical solutions for sloping bays \cite{Synolakis08,Kanoglu15}. With local bathymetry significantly effecting the run-up of tsunami waves, the proposed analytical solution, along with other existing analytical solutions \cite{Zahibo05,Rybkin,Anderson}, allow verification of tsunami models in a more realistic settings of a 2-D bathymetry. Furthermore, 1-D shallow water theory has been shown to have similar inundation predictions to the full 2-D models in the realistic setting of Alaskan fjords, with significantly less computation time \cite{Harris}. Incorporating 1-D shallow water theory into large scale tsunami inundation models may significantly reduce computation and forecasting time \cite{Harris,Anderson}, potentially saving lives and resources.

When modeling tsunamis generated by near shore earthquakes and submarine landslides, initial conditions must be prescribed directly on the sloping beach. The previous IVP solutions had some limitations to modeling such phenomena because of the linearization of the initial conditions \cite{Yeh03} or certain implicit assumptions \cite{Kanoglu06}. The presented solution is a further step towards modeling such near-field events, which are the Achilles heel of current tsunami models \cite{Kanoglu15,Synolakis06,TsunamiDynamics15}. The initial conditions associated with submarine landslides are still debated and we refer the interested reader to \cite{Lovholt15}.

On the other hand, earthquakes in the open ocean generate waves that propagate from far off-shore, and then deform over a sloping beach. This problem is modeled as a boundary value problem \cite{Kanoglu06,Synolakis87}, though specifying the boundary condition is nontrivial \cite{Antuono07}. Analysis of our solution infers that the zeroth order projection of the initial conditions (i.e. $\Phi_0$) is likely adequate for modeling runup of geophysical tsunamis. As the ocean side of the sloping beach is usually far from shore, the wave should behave linearly at this boundary, and zeroth order approximation of the boundary condition could be sufficient. To model water velocities is likely necessary to use the first order projection (i.e. $\Phi_1$), however further investigations are necessary. We also would like to emphasize the computational expenses to compute various projections $\Phi_n$, $n=0,\dots,10$ according to (\ref{ref:iterative_phi}) are negligibly small in comparison to evaluating of integral in (\ref{eq:Phi}).

To conclude, in this letter we formulate a new complete solution to the IVP of the cross-sectionally averaged SWE for initial conditions with and without initial velocity. This proposed solution deals with the difficulty associated with the initial conditions given on a curve in the transformed space, an important subtlety not previously acknowledged, and also avoids linearizion of the spatial coordinate in the transformation of the initial conditions. This allows modeling problems with near shore initial conditions, and extends earlier solutions beyond waves with small initial velocities. It also extends the solution from only plane beaches to more complex U-shaped bathymetries. Our proposed solution can be used for analytical verification of tsunami models in realistic 2-D settings, and may potentially allow fast tsunami forecasting in narrow bays and fjords.

\appendix*
\section{Appendix: Analysis of solution in \cite{Kanoglu06}}
We would like to examine a solution to the non-linear shallow water equation provided in \cite{Kanoglu06}. In particular, we would like to analyze formula (4) in \cite{Kanoglu06}, namely:
\begin{equation}
\psi(\sigma,\lambda)=2\int_0^\infty\Big[\psi(\xi,\lambda_0)G_\lambda+\psi_\lambda(\xi,\lambda_0)G\Big]d\xi. \label{solution formula}
\end{equation}
Here, the quantity $G$ is given by
\[
G(\xi,\sigma,\bar\lambda)=\xi\int_0^\infty{J_0}(\omega\xi)J_0(\omega\sigma)\sin\big(\frac12\omega\bar\lambda\big)d\omega,
\]
where $\bar\lambda=\lambda-\lambda_0$. Note that $\lambda_0=\lambda_0(\sigma)=-u_0(x(\sigma))$. We now show that if $\psi(\sigma,\lambda)$ is given by Eq. (\ref{solution formula}) then
\begin{equation}
4\sigma\psi_{\lambda\lambda}-(\sigma\psi_{\sigma})_{\sigma}=O(\frac{\partial{\lambda_{0}}}{\partial\sigma})\not =0\label{eq4}
\end{equation}
which means that (\ref{solution formula}) is not an exact solution to the shallow water equations in the transformed coordinates, see formula (3) in \cite{Kanoglu06}.

To this end, consider
\[
I(\sigma,\lambda)\equiv2\int_{0}^{\infty}\Big[\psi(\xi,\lambda_{0})G_{\lambda}+\psi_{\lambda}(\xi,\lambda_{0})G\Big]d\xi.
\]
Inserting the expression for $G$ (the Green's function) in the integral, one
has%
\begin{align*}
I  & =\int_{0}^{\infty}\psi(\xi,\lambda_{0})\Big[\xi\int_{0}^{\infty}\omega{J_{0}}(\omega\xi)J_{0}(\omega\sigma)\cos\big(\frac{1}{2}\omega\bar{\lambda}\big)d\omega\Big]d\xi\\
& +2\int_{0}^{\infty}\psi_{\lambda}(\xi,\lambda_{0})\Big[\xi\int_{0}^{\infty}{J_{0}}(\omega\xi)J_{0}(\omega\sigma)\sin\big(\frac{1}{2}\omega\bar{\lambda}\big)d\omega\Big]d\xi.
\end{align*}
Interchanging the order of integration yields
\begin{align*}
I  & =\int_{0}^{\infty}\omega{J_{0}}(\omega\sigma)\cos\big(\frac{1}{2}\omega\bar{\lambda}\big)\int_{0}^{\infty}\xi{J_{0}}(\omega\xi)\psi(\xi,\lambda_{0})d\xi d\omega\\
& +2\int_{0}^{\infty}{J_{0}}(\omega\sigma)\sin\big(\frac{1}{2}\omega\bar{\lambda}\big)\int_{0}^{\infty}\xi{J_{0}}(\omega\xi)\psi_{\lambda}(\xi,\lambda_{0})d\omega d\xi\\
& =\int_{0}^{\infty}\omega{J_{0}}(\omega\sigma)\cos\big(\frac{1}{2}\omega\bar{\lambda}\big)\overline{P}(\omega)d\omega+2\int_{0}^{\infty}{J_{0}}(\omega\sigma)\sin\big(\frac{1}{2}\omega\bar{\lambda}\big)\overline{F}(\omega)d\omega\\
& =I_{1}+I_{2},
\end{align*}
where
\[
\overline{P}(\omega)=\int_{0}^{\infty}\sigma{J_{0}}(\omega\sigma)\psi(\sigma,\lambda_{0})d\sigma,~~~~~~~~\overline{F}(\omega)=\int_{0}^{\infty}\sigma{J_{0}}(\omega\sigma)\psi_{\lambda}(\sigma,\lambda_{0})d\sigma
\]
are the Fourier-Hankel transforms of $\psi(\sigma,\lambda_{0})$ and
$\psi_{\lambda}(\sigma,\lambda_{0})$.

We now substitute $I$ in the left hand side of equation (\ref{eq4}). Since
(\ref{eq4}) is linear and $\overline{P}(\omega)$ and $\overline{F}(\omega)$
are independent, it is enough to consider only
\[
I_{1}(\sigma,\lambda)=\int_{0}^{\infty}\omega{J_{0}}(\omega\sigma)\cos(\frac{1}{2}\omega\bar{\lambda})\overline{P}(\omega)d\omega.
\]
By a straightforward computation we obtain
\[
4\sigma\frac{\partial^{2}}{\partial\lambda^{2}}{I_{1}}-\frac{\partial}{\partial\sigma}\Big(\sigma\frac{\partial}{\partial\sigma}{I_{1}}\Big)=\int_{0}^{\infty}\omega\Delta_{1}(\sigma,\lambda,\omega)\overline{P}(\omega)d\omega,
\]
where
\begin{align*}
&  \Delta_{1}(\sigma,\lambda,\omega)\\
&  =\sigma(-\omega^{2}){J_{0}}(\omega\sigma)\cos(\frac{1}{2}\omega\bar{\lambda})-\frac{\partial}{\partial\sigma}\Big(\sigma\frac{\partial}{\partial\sigma}\big({J_{0}}(\omega\sigma)\cos(\frac{1}{2}\omega\bar{\lambda})\big)\Big)\\
&  =-\frac{\omega}{2}\Big(2\sigma\frac{\partial{J_{0}(\omega\sigma)}}{\partial\sigma}+{J_{0}}(\omega\sigma)\Big)\sin(\frac{1}{2}\omega\bar{\lambda})\frac{\partial{\lambda_{0}}}{\partial\sigma}\\
&  +\frac{\omega}{2}\sigma{J_{0}}(\omega\sigma)\frac{\partial}{\partial\sigma}\Big(\sin(\frac{1}{2}\omega\bar{\lambda})\frac{\partial\lambda_{0}}{\partial\sigma}\Big)\\
&  =O(\frac{\partial{\lambda_{0}}}{\partial\sigma}).
\end{align*}
Consequently,
\[
4\sigma\frac{\partial^{2}}{\partial\lambda^{2}}{I}-\frac{\partial}{\partial\sigma}\Big(\sigma\frac{\partial}{\partial\sigma}{I}\Big)=O(\frac{\partial{\lambda_{0}}}{\partial\sigma})
\]
and no better. This implies that formula (\ref{solution formula}) does not
solve the wave equation (\ref{eq4}), but provides an approximation to the
solution up to order at least ${\partial{\lambda_{0}}}/{\partial\sigma}$ (or
equivalently ${\partial{u_{0}}}/{\partial\sigma}$). We note that if the
initial velocity $u_{0}=0$ then ${{\lambda_{0}=0}}$ and the error term
vanishes. I.e. $I\left(  \sigma,\lambda\right)  $ is then indeed an exact
solution (but not in general).
\vspace{0.1in}

\begin{acknowledgments}
A. Raz was supported by the National Science Foundation Research Experience for Undergraduate program (Grant \# 1411560) and the Geophysical Institute, University of Alaska Fairbanks. D. Nicolsky acknowledges support from the Geophysical Institute, University of Alaska Fairbanks. A. Rybkin acknowledges support from National Science Foundation Grant DMS-1411560. E. Pelinovsky acknowledges grants from RF Ministry of Education and Science (project No. 5.5176.2017/8.9), RF President program NS-6637.2016.5 and RFBR (17-05-00067).
\end{acknowledgments}


\begin{thebibliography}{24}%
\makeatletter
\providecommand \@ifxundefined [1]{%
 \@ifx{#1\undefined}
}%
\providecommand \@ifnum [1]{%
 \ifnum #1\expandafter \@firstoftwo
 \else \expandafter \@secondoftwo
 \fi
}%
\providecommand \@ifx [1]{%
 \ifx #1\expandafter \@firstoftwo
 \else \expandafter \@secondoftwo
 \fi
}%
\providecommand \natexlab [1]{#1}%
\providecommand \enquote  [1]{``#1''}%
\providecommand \bibnamefont  [1]{#1}%
\providecommand \bibfnamefont [1]{#1}%
\providecommand \citenamefont [1]{#1}%
\providecommand \href@noop [0]{\@secondoftwo}%
\providecommand \href [0]{\begingroup \@sanitize@url \@href}%
\providecommand \@href[1]{\@@startlink{#1}\@@href}%
\providecommand \@@href[1]{\endgroup#1\@@endlink}%
\providecommand \@sanitize@url [0]{\catcode `\\12\catcode `\$12\catcode
  `\&12\catcode `\#12\catcode `\^12\catcode `\_12\catcode `\%12\relax}%
\providecommand \@@startlink[1]{}%
\providecommand \@@endlink[0]{}%
\providecommand \url  [0]{\begingroup\@sanitize@url \@url }%
\providecommand \@url [1]{\endgroup\@href {#1}{\urlprefix }}%
\providecommand \urlprefix  [0]{URL }%
\providecommand \Eprint [0]{\href }%
\providecommand \doibase [0]{http://dx.doi.org/}%
\providecommand \selectlanguage [0]{\@gobble}%
\providecommand \bibinfo  [0]{\@secondoftwo}%
\providecommand \bibfield  [0]{\@secondoftwo}%
\providecommand \translation [1]{[#1]}%
\providecommand \BibitemOpen [0]{}%
\providecommand \bibitemStop [0]{}%
\providecommand \bibitemNoStop [0]{.\EOS\space}%
\providecommand \EOS [0]{\spacefactor3000\relax}%
\providecommand \BibitemShut  [1]{\csname bibitem#1\endcsname}%
\let\auto@bib@innerbib\@empty
\bibitem [{\citenamefont {Kanoglu}\ \emph {et~al.}(2015)\citenamefont
  {Kanoglu}, \citenamefont {Titov}, \citenamefont {Bernard},\ and\
  \citenamefont {Synolakis}}]{Kanoglu15}%
  \BibitemOpen
  \bibfield  {author} {\bibinfo {author} {\bibfnamefont {U.}~\bibnamefont
  {Kanoglu}}, \bibinfo {author} {\bibfnamefont {V.}~\bibnamefont {Titov}},
  \bibinfo {author} {\bibfnamefont {E.}~\bibnamefont {Bernard}}, \ and\
  \bibinfo {author} {\bibfnamefont {C.}~\bibnamefont {Synolakis}},\ }\href
  {\doibase 10.1098/rsta.2014.0369} {\bibfield  {journal} {\bibinfo  {journal}
  {Philosophical Transactions of the Royal Society A}\ }\textbf {\bibinfo
  {volume} {373 (2053)}},\ \bibinfo {pages} {20140369} (\bibinfo {year}
  {2015})}\BibitemShut {NoStop}%
\bibitem [{\citenamefont {NTHMP}(2012)}]{NTHMP12}%
  \BibitemOpen
  \bibinfo {editor} {\bibnamefont {NTHMP}},\ ed.,\ \href@noop {} {\emph
  {\bibinfo {title} {Proceedings and results of the 2011 NTHMP Model
  Benchmarking Workshop}}},\ NOAA Special Report,\ \bibinfo {organization}
  {U.S. Department of Commerce/NOAA/NTHMP}\ (\bibinfo  {publisher} {National
  Tsunami Hazard Mapping Program [NTHMP]},\ \bibinfo {address} {Boulder, CO},\
  \bibinfo {year} {2012})\ \bibinfo {note} {436 p.}\BibitemShut {Stop}%
\bibitem [{\citenamefont {Synolakis}\ \emph {et~al.}(2008)\citenamefont
  {Synolakis}, \citenamefont {Bernard}, \citenamefont {Titov}, \citenamefont
  {Kanoglu},\ and\ \citenamefont {Gonzalez}}]{Synolakis08}%
  \BibitemOpen
  \bibfield  {author} {\bibinfo {author} {\bibfnamefont {C.}~\bibnamefont
  {Synolakis}}, \bibinfo {author} {\bibfnamefont {E.}~\bibnamefont {Bernard}},
  \bibinfo {author} {\bibfnamefont {V.}~\bibnamefont {Titov}}, \bibinfo
  {author} {\bibfnamefont {U.}~\bibnamefont {Kanoglu}}, \ and\ \bibinfo
  {author} {\bibfnamefont {F.}~\bibnamefont {Gonzalez}},\ }\href@noop {}
  {\bibfield  {journal} {\bibinfo  {journal} {Pure Applied Geophysics}\
  }\textbf {\bibinfo {volume} {165}},\ \bibinfo {pages} {2197} (\bibinfo {year}
  {2008})}\BibitemShut {NoStop}%
\bibitem [{\citenamefont {Kanoglu}\ and\ \citenamefont
  {Synolakis}(2006)}]{Kanoglu06}%
  \BibitemOpen
  \bibfield  {author} {\bibinfo {author} {\bibfnamefont {U.}~\bibnamefont
  {Kanoglu}}\ and\ \bibinfo {author} {\bibfnamefont {C.}~\bibnamefont
  {Synolakis}},\ }\href {\doibase 10.1103/PhysRevLett.97.148501} {\bibfield
  {journal} {\bibinfo  {journal} {Physical Review Letters}\ }\textbf {\bibinfo
  {volume} {148501}},\ \bibinfo {pages} {97} (\bibinfo {year}
  {2006})}\BibitemShut {NoStop}%
\bibitem [{\citenamefont {Synolakis}\ and\ \citenamefont
  {Bernard}(2006)}]{Synolakis06}%
  \BibitemOpen
  \bibfield  {author} {\bibinfo {author} {\bibfnamefont {C.}~\bibnamefont
  {Synolakis}}\ and\ \bibinfo {author} {\bibfnamefont {E.}~\bibnamefont
  {Bernard}},\ }\href@noop {} {\bibfield  {journal} {\bibinfo  {journal}
  {Philosophical Transactions of the Royal Society A}\ }\textbf {\bibinfo
  {volume} {364}},\ \bibinfo {pages} {2231} (\bibinfo {year}
  {2006})}\BibitemShut {NoStop}%
\bibitem [{\citenamefont {Kanoglu}\ and\ \citenamefont
  {Synolakis}(2015)}]{TsunamiDynamics15}%
  \BibitemOpen
  \bibfield  {author} {\bibinfo {author} {\bibfnamefont {U.}~\bibnamefont
  {Kanoglu}}\ and\ \bibinfo {author} {\bibfnamefont {C.}~\bibnamefont
  {Synolakis}},\ }\enquote {\bibinfo {title} {Coastal and marine hazards,
  risks, and disasters},}\ \ (\bibinfo  {publisher} {ELSEVIER},\ \bibinfo
  {year} {2015})\ Chap.\ \bibinfo {chapter} {Tsunami Dynamics, Forecasting, and
  Mitigation}, pp.\ \bibinfo {pages} {15--57}\BibitemShut {NoStop}%
\bibitem [{\citenamefont {Synolakis}(1987)}]{Synolakis87}%
  \BibitemOpen
  \bibfield  {author} {\bibinfo {author} {\bibfnamefont {C.}~\bibnamefont
  {Synolakis}},\ }\href@noop {} {\bibfield  {journal} {\bibinfo  {journal} {J.
  Fluid Mech.}\ }\textbf {\bibinfo {volume} {185}},\ \bibinfo {pages} {523}
  (\bibinfo {year} {1987})}\BibitemShut {NoStop}%
\bibitem [{\citenamefont {Antuono}\ and\ \citenamefont
  {Brocchini}(2010)}]{Antuono10}%
  \BibitemOpen
  \bibfield  {author} {\bibinfo {author} {\bibfnamefont {M.}~\bibnamefont
  {Antuono}}\ and\ \bibinfo {author} {\bibfnamefont {M.}~\bibnamefont
  {Brocchini}},\ }\href@noop {} {\bibfield  {journal} {\bibinfo  {journal}
  {Journal of Fluid Mechanics}\ }\textbf {\bibinfo {volume} {643}},\ \bibinfo
  {pages} {207} (\bibinfo {year} {2010})}\BibitemShut {NoStop}%
\bibitem [{\citenamefont {Carrier}\ and\ \citenamefont
  {Greenspan}(1957)}]{Carrier1}%
  \BibitemOpen
  \bibfield  {author} {\bibinfo {author} {\bibfnamefont {G.}~\bibnamefont
  {Carrier}}\ and\ \bibinfo {author} {\bibfnamefont {H.}~\bibnamefont
  {Greenspan}},\ }\href@noop {} {\bibfield  {journal} {\bibinfo  {journal} {J.
  Fluid Mech.}\ }\textbf {\bibinfo {volume} {01}},\ \bibinfo {pages} {97}
  (\bibinfo {year} {1957})}\BibitemShut {NoStop}%
\bibitem [{\citenamefont {Carrier}\ \emph {et~al.}(2003)\citenamefont
  {Carrier}, \citenamefont {Wu},\ and\ \citenamefont {Yeh}}]{Yeh03}%
  \BibitemOpen
  \bibfield  {author} {\bibinfo {author} {\bibfnamefont {G.}~\bibnamefont
  {Carrier}}, \bibinfo {author} {\bibfnamefont {T.}~\bibnamefont {Wu}}, \ and\
  \bibinfo {author} {\bibfnamefont {H.}~\bibnamefont {Yeh}},\ }\href {\doibase
  10.1017/S0022112002002653} {\bibfield  {journal} {\bibinfo  {journal} {J.
  Fluid Mech.}\ }\textbf {\bibinfo {volume} {475}},\ \bibinfo {pages} {79}
  (\bibinfo {year} {2003})}\BibitemShut {NoStop}%
\bibitem [{\citenamefont {Kanoglu}(2004)}]{KANOGLU04}%
  \BibitemOpen
  \bibfield  {author} {\bibinfo {author} {\bibfnamefont {U.}~\bibnamefont
  {Kanoglu}},\ }\href@noop {} {\bibfield  {journal} {\bibinfo  {journal} {J.
  Fluid Mech.}\ }\textbf {\bibinfo {volume} {513}},\ \bibinfo {pages} {363}
  (\bibinfo {year} {2004})}\BibitemShut {NoStop}%
\bibitem [{\citenamefont {Harris}\ \emph
  {et~al.}(2015{\natexlab{a}})\citenamefont {Harris}, \citenamefont {Nicolsky},
  \citenamefont {Pelinovsky}, \citenamefont {Pender},\ and\ \citenamefont
  {Rybkin}}]{Harris15}%
  \BibitemOpen
  \bibfield  {author} {\bibinfo {author} {\bibfnamefont {M.}~\bibnamefont
  {Harris}}, \bibinfo {author} {\bibfnamefont {D.}~\bibnamefont {Nicolsky}},
  \bibinfo {author} {\bibfnamefont {E.}~\bibnamefont {Pelinovsky}}, \bibinfo
  {author} {\bibfnamefont {J.}~\bibnamefont {Pender}}, \ and\ \bibinfo {author}
  {\bibfnamefont {A.}~\bibnamefont {Rybkin}},\ }\href@noop {} {\bibfield
  {journal} {\bibinfo  {journal} {J. Ocean Eng. Mar. Energy}\ } (\bibinfo
  {year} {2015}{\natexlab{a}})}\BibitemShut {NoStop}%
\bibitem [{\citenamefont {Aydin}\ and\ \citenamefont
  {Kanoglu}(2017)}]{Aydin17}%
  \BibitemOpen
  \bibfield  {author} {\bibinfo {author} {\bibfnamefont {B.}~\bibnamefont
  {Aydin}}\ and\ \bibinfo {author} {\bibfnamefont {U.}~\bibnamefont
  {Kanoglu}},\ }\href@noop {} {\bibfield  {journal} {\bibinfo  {journal} {Pure
  and Applied Geophysics}\ }\textbf {\bibinfo {volume} {174}},\ \bibinfo
  {pages} {3209} (\bibinfo {year} {2017})}\BibitemShut {NoStop}%
\bibitem [{\citenamefont {Antuono}\ and\ \citenamefont
  {Brocchini}(2007)}]{Antuono07}%
  \BibitemOpen
  \bibfield  {author} {\bibinfo {author} {\bibfnamefont {M.}~\bibnamefont
  {Antuono}}\ and\ \bibinfo {author} {\bibfnamefont {M.}~\bibnamefont
  {Brocchini}},\ }\href@noop {} {\bibfield  {journal} {\bibinfo  {journal}
  {Studies in Applied Mathematics}\ }\textbf {\bibinfo {volume} {119}},\
  \bibinfo {pages} {73} (\bibinfo {year} {2007})}\BibitemShut {NoStop}%
\bibitem [{\citenamefont {Tuck}\ and\ \citenamefont {Hwang}(1972)}]{Tuck72}%
  \BibitemOpen
  \bibfield  {author} {\bibinfo {author} {\bibfnamefont {E.}~\bibnamefont
  {Tuck}}\ and\ \bibinfo {author} {\bibfnamefont {L.}~\bibnamefont {Hwang}},\
  }\href@noop {} {\bibfield  {journal} {\bibinfo  {journal} {J. Fluid Mech.}\
  }\textbf {\bibinfo {volume} {51}},\ \bibinfo {pages} {449–461} (\bibinfo
  {year} {1972})}\BibitemShut {NoStop}%
\bibitem [{\citenamefont {Zahibo}\ \emph {et~al.}(2006)\citenamefont {Zahibo},
  \citenamefont {Pelinovsky}, \citenamefont {Golinko},\ and\ \citenamefont
  {Osipenko}}]{Zahibo05}%
  \BibitemOpen
  \bibfield  {author} {\bibinfo {author} {\bibfnamefont {N.}~\bibnamefont
  {Zahibo}}, \bibinfo {author} {\bibfnamefont {E.}~\bibnamefont {Pelinovsky}},
  \bibinfo {author} {\bibfnamefont {V.}~\bibnamefont {Golinko}}, \ and\
  \bibinfo {author} {\bibfnamefont {N.}~\bibnamefont {Osipenko}},\ }\href@noop
  {} {\bibfield  {journal} {\bibinfo  {journal} {International Journal of Fluid
  Mechanics Research 33.1}\ ,\ \bibinfo {pages} {106}} (\bibinfo {year}
  {2006})}\BibitemShut {NoStop}%
\bibitem [{\citenamefont {Rybkin}\ \emph {et~al.}(2014)\citenamefont {Rybkin},
  \citenamefont {Pelinovsky},\ and\ \citenamefont {Didenkulova}}]{Rybkin}%
  \BibitemOpen
  \bibfield  {author} {\bibinfo {author} {\bibfnamefont {A.}~\bibnamefont
  {Rybkin}}, \bibinfo {author} {\bibfnamefont {E.}~\bibnamefont {Pelinovsky}},
  \ and\ \bibinfo {author} {\bibfnamefont {I.}~\bibnamefont {Didenkulova}},\
  }\href@noop {} {\bibfield  {journal} {\bibinfo  {journal} {J. Fluid Mech.}\
  }\textbf {\bibinfo {volume} {748}},\ \bibinfo {pages} {416} (\bibinfo {year}
  {2014})}\BibitemShut {NoStop}%
\bibitem [{\citenamefont {Anderson}\ \emph {et~al.}(2017)\citenamefont
  {Anderson}, \citenamefont {Harris}, \citenamefont {Hartle}, \citenamefont
  {Nicolsky}, \citenamefont {Pelinovsky}, \citenamefont {Raz},\ and\
  \citenamefont {Rybkin}}]{Anderson}%
  \BibitemOpen
  \bibfield  {author} {\bibinfo {author} {\bibfnamefont {D.}~\bibnamefont
  {Anderson}}, \bibinfo {author} {\bibfnamefont {M.}~\bibnamefont {Harris}},
  \bibinfo {author} {\bibfnamefont {H.}~\bibnamefont {Hartle}}, \bibinfo
  {author} {\bibfnamefont {D.}~\bibnamefont {Nicolsky}}, \bibinfo {author}
  {\bibfnamefont {E.}~\bibnamefont {Pelinovsky}}, \bibinfo {author}
  {\bibfnamefont {A.}~\bibnamefont {Raz}}, \ and\ \bibinfo {author}
  {\bibfnamefont {A.}~\bibnamefont {Rybkin}},\ }\href {\doibase
  10.1007/s00024-017-1476-3} {\bibfield  {journal} {\bibinfo  {journal}
  {Journal of Pure and Applied Geophysics}\ } (\bibinfo {year} {2017}),\
  10.1007/s00024-017-1476-3}\BibitemShut {NoStop}%
\bibitem [{\citenamefont {Shimozono}\ and\ \citenamefont
  {Takenori}(2016)}]{Shimozono16}%
  \BibitemOpen
  \bibfield  {author} {\bibinfo {author} {\bibnamefont {Shimozono}}\ and\
  \bibinfo {author} {\bibnamefont {Takenori}},\ }\href {\doibase
  10.1017/jfm.2016.327} {\bibfield  {journal} {\bibinfo  {journal} {Journal of
  Fluid Mechanics}\ }\textbf {\bibinfo {volume} {798}},\ \bibinfo {pages} {457}
  (\bibinfo {year} {2016})}\BibitemShut {NoStop}%
\bibitem [{\citenamefont {Madsen}\ \emph {et~al.}(2008)\citenamefont {Madsen},
  \citenamefont {Fuhrman},\ and\ \citenamefont {Sch{\"a}ffer}}]{MadsenFuhrman}%
  \BibitemOpen
  \bibfield  {author} {\bibinfo {author} {\bibfnamefont {P.}~\bibnamefont
  {Madsen}}, \bibinfo {author} {\bibfnamefont {D.}~\bibnamefont {Fuhrman}}, \
  and\ \bibinfo {author} {\bibfnamefont {H.}~\bibnamefont {Sch{\"a}ffer}},\
  }\href@noop {} {\bibfield  {journal} {\bibinfo  {journal} {Journal of
  Geophysical Research; Oceans}\ }\textbf {\bibinfo {volume} {113}} (\bibinfo
  {year} {2008})}\BibitemShut {NoStop}%
\bibitem [{\citenamefont {Synolakis}\ \emph {et~al.}(1997)\citenamefont
  {Synolakis}, \citenamefont {Liu}, \citenamefont {Philip}, \citenamefont
  {Carrier},\ and\ \citenamefont {Yeh}}]{Synolokis97}%
  \BibitemOpen
  \bibfield  {author} {\bibinfo {author} {\bibfnamefont {C.~E.}\ \bibnamefont
  {Synolakis}}, \bibinfo {author} {\bibfnamefont {P.}~\bibnamefont {Liu}},
  \bibinfo {author} {\bibfnamefont {H.~A.}\ \bibnamefont {Philip}}, \bibinfo
  {author} {\bibfnamefont {G.}~\bibnamefont {Carrier}}, \ and\ \bibinfo
  {author} {\bibfnamefont {H.}~\bibnamefont {Yeh}},\ }\href@noop {} {\bibfield
  {journal} {\bibinfo  {journal} {Science}\ }\textbf {\bibinfo {volume}
  {278}},\ \bibinfo {pages} {598} (\bibinfo {year} {1997})}\BibitemShut
  {NoStop}%
\bibitem [{\citenamefont {Bosa}\ and\ \citenamefont {Petti}(2011)}]{Vajont11}%
  \BibitemOpen
  \bibfield  {author} {\bibinfo {author} {\bibfnamefont {S.}~\bibnamefont
  {Bosa}}\ and\ \bibinfo {author} {\bibfnamefont {M.}~\bibnamefont {Petti}},\
  }\href {\doibase 10.1016/j.envsoft.2010.10.001} {\bibfield  {journal}
  {\bibinfo  {journal} {Environmental Modelling and Software}\ }\textbf
  {\bibinfo {volume} {26}},\ \bibinfo {pages} {406} (\bibinfo {year}
  {2011})}\BibitemShut {NoStop}%
\bibitem [{\citenamefont {Harris}\ \emph
  {et~al.}(2015{\natexlab{b}})\citenamefont {Harris}, \citenamefont {Nicolsky},
  \citenamefont {Pelinovsky},\ and\ \citenamefont {Rybkin}}]{Harris}%
  \BibitemOpen
  \bibfield  {author} {\bibinfo {author} {\bibfnamefont {M.}~\bibnamefont
  {Harris}}, \bibinfo {author} {\bibfnamefont {D.}~\bibnamefont {Nicolsky}},
  \bibinfo {author} {\bibfnamefont {E.}~\bibnamefont {Pelinovsky}}, \ and\
  \bibinfo {author} {\bibfnamefont {A.}~\bibnamefont {Rybkin}},\ }\href@noop {}
  {\bibfield  {journal} {\bibinfo  {journal} {Pure and Applied Geophysics}\
  }\textbf {\bibinfo {volume} {172}},\ \bibinfo {pages} {885} (\bibinfo {year}
  {2015}{\natexlab{b}})},\ \bibinfo {note} {doi:
  10.1007/s00024-014-1016-3}\BibitemShut {NoStop}%
\bibitem [{\citenamefont {L{\o}vholt}\ \emph {et~al.}(2015)\citenamefont
  {L{\o}vholt}, \citenamefont {Pedersen}, \citenamefont {Harbitz},
  \citenamefont {Glimsdal},\ and\ \citenamefont {Kim}}]{Lovholt15}%
  \BibitemOpen
  \bibfield  {author} {\bibinfo {author} {\bibfnamefont {F.}~\bibnamefont
  {L{\o}vholt}}, \bibinfo {author} {\bibfnamefont {G.}~\bibnamefont
  {Pedersen}}, \bibinfo {author} {\bibfnamefont {C.~B.}\ \bibnamefont
  {Harbitz}}, \bibinfo {author} {\bibfnamefont {S.}~\bibnamefont {Glimsdal}}, \
  and\ \bibinfo {author} {\bibfnamefont {J.}~\bibnamefont {Kim}},\ }\href
  {\doibase 10.1098/rsta.2014.0376} {\bibfield  {journal} {\bibinfo  {journal}
  {Philosophical Transactions of the Royal Society of London A: Mathematical,
  Physical and Engineering Sciences}\ }\textbf {\bibinfo {volume} {373}}
  (\bibinfo {year} {2015}),\ 10.1098/rsta.2014.0376}\BibitemShut {NoStop}%
\end{thebibliography}

%

\end{document}